\documentclass[aps,pra,showpacs,twocolumn,nofootinbib]{revtex4}

\usepackage{dcolumn}
\usepackage{amsmath}

\begin{document}

\title{Energy levels and lifetimes of Nd~IV, Pm~IV, Sm~IV, and Eu~IV }

\author{V. A. Dzuba}
\email{dzuba@newt.phys.unsw.edu.au} \affiliation{School of
Physics, University of New South Wales,\\ Sydney 2052, Australia}
\author{U. I. Safronova}
\email{usafrono@nd.edu} \affiliation{
Department of Physics, 225 Nieuwland Science Hall\\
University of Notre Dame, Notre Dame, Indiana 46556}
\author{W. R. Johnson}
\email{johnson@nd.edu} \homepage{www.nd.edu/~johnson}
\affiliation {Department of Physics, 225 Nieuwland Science Hall\\
University of Notre Dame, Notre Dame, Indiana 46556}

\date{\today}

\begin{abstract}

To address the shortage of experimental data for electron spectra
of triply-ionized rare earth elements we have calculated energy levels 
and lifetimes of $4f^{n+1}$ and $4f^{n}5d$ configurations of Nd~IV ($n$=2), 
Pm~IV ($n$=3), Sm~IV ($n$=4), and Eu~IV ($n$=5) using Hartree-Fock and 
configuration interaction methods. To control the accuracy of our 
calculations we also performed similar calculations for Pr~III, Nd~III
and Sm~III, for which experimental data are available. 
The results are important, in particular, for physics of magnetic
garnets.

\end{abstract}


\maketitle

\section{introduction}

Magnetic properties of garnets are mostly due to unpaired 4f-electrons
of the rare earth elements which are present in the garnets as triply-ionized
ions (see, e.g. \cite{Garnets}). Our interest in magnetic garnets was initially
driven by proposals to use gadolinium gallium garnet and gadolinium iron
garnet to measure the electron electric dipole moment (EDM) \cite{lam,Hun}.
In our earlier works we performed calculations to relate the electron EDM
to measurable values \cite{Kuen,Buh}. It turned out, in particular, that the
value of the EDM enhancement factor is sensitive to the energy interval 
between the ground
state configuration $4f^7$ and excited configuration $4f^65d$ of Gd$^{3+}$.
The lack of experimental data on the Gd$^{3+}$ spectra produced an initially
large uncertainty which we were able to overcome by performing {\em ab initio}
calculations of the Gd$^{3+}$ energy levels \cite{dzuba-gd}. 

While gadolinium garnets seem to be the best choice for the EDM measurements
\cite{lam,Hun}, other garnets may also be useful for the measurements of the 
T,P-odd effects (EDM, nuclear anapole moment, etc.). 
Therefore, knowing the spectra of triply-ionized rare earth elements would be 
useful for future analysis. Unfortunately, the corresponding experimental data 
is very poor. In the present paper we calculate energy levels of the two most 
important configurations $4f^{n+1}$ and $4f^{n}5d$ for all rare-earth atoms 
for which corresponding experimental data is not available. This includes 
Nd~IV ($n$=2), Pm~IV ($n$=3), Sm~IV ($n$=4) and Eu~IV ($n$=5), but excludes 
Gd~IV for which calculations were done previously \cite{dzuba-gd}.

In the present
paper, as in the earlier one \cite{dzuba-gd}, we consider free ions only.
Effects of the garnet environment on the 
energy levels of rare earth ions were considered by 
\citet{Dorenbos} and can probably be used to extrapolate the 
spectra obtained here to the garnet environment. 
In any case, the spectra of free rare earth ions have 
their own significance.

We also perform calculations for doubly-ionized ions of
rare earth elements Pr, Nd and Sm, which have the same
electronic structure as triply ionized Nd, Pm and Eu, respectively. 
Since experimental spectra are available for the Pr~III, Nd~III and Sm~III
comparison of theoretical energies with experiment allowed us to control
the accuracy of our calculations. 

A comprehensive set of theoretical data for the $4f^{n+1}$ -
$4f^{n}5d$ transitions including comparisons with
data recommended by the National Institute of 
Standards and Technology (NIST) \cite{martin} 
is given in Tables~I-IX of the accompanying
EPAPS document \cite{EPAPS}.

\section{Calculation of energies}
\label{energies}

Calculations in this paper are similar to those of our
previous paper \cite{dzuba-gd}. We use our relativistic Hartree-Fock
and configuration interaction (RCI) codes as well as 
a set of computer codes written by \citet{book} and freely available
via the Internet \cite{code}. 

We start our calculations from the relativistic Hartree-Fock method. 
Since we are dealing with open-shell systems with a fractionally occupied
$4f$ shell, the Hartree-Fock procedure needs to be further specified.
We perform calculations in the $V^N$ approximation, which means that all
atomic electrons contribute to the Hartree-Fock potential. The contribution
of the fully-occupied $4f$ shell is
multiplied by a weighting factor $q/2(2l+1)$ were $q$ is the actual 
occupation number and $l=3$ is the angular momentum. When the $4f$
orbital itself is calculated, 
the weighting factor is further reduced to $(q-1)/(2(2l+1)-1)$.
Although the calculations are relativistic and the 
$4f_{5/2}$ and $4f_{7/2}$ states
are different, we use the same weighting factor for both states.
We also need the $5d_{3/2}$ and $5d_{5/2}$ states for the $4f^{n-1}5d$
configuration. These states are calculated in the field of a frozen core
from which one $4f$ electron is removed. This corresponds to the weighting
factor $(q-1)/2(2l+1)$ for the contribution of the $4f$ electrons to the
Hartree-Fock potential.

We now have four basis states $4f_{5/2},\, 4f_{7/2},\, 5d_{3/2}$ and $5d_{5/2}$
for the configuration interaction (CI) calculations. A calculation with
these basis states and the standard CI procedure underestimates 
the energy interval between the $4f^n$ and $4f^{n-1}5d$ configurations.
As shown by \citet{igor}, this underestimate is due to the
 core polarization by the field of external electrons.
In Ref.~\cite{igor}, correlation corrections to the energy of an 
external electron 
in Ce~IV or Pr~V in the $4f$ or $5d$ state were calculated to third order 
in many-body perturbation theory. 
It was found that  core polarization increased the $4f$-$5d$ energy
interval by about 20000~cm$^{-1}$, bringing theoretical energies into very good
agreement with the experiment. Similar energy shift of about 18000~cm$^{-1}$
between the $4f^7$ and $4f^65d$ configurations of Gd~IV were introduced 
 ``by hand'' into
the calculations with the Cowan code \cite{dzuba-gd}.

We use a somewhat different approach in our RCI calculations, 
where we simulate core polarization by 
adding a polarization correction 
\begin{equation}
        \delta V = - \frac{\alpha}{2(a^4+r^4)},
\label{deltaV}
\end{equation}
to the HF potential in which the $5d$ state is calculated.
Here $\alpha$ is polarizability of an ion in the $4f^n$
configuration, $a$ is a cut-off parameter introduced to remove the
singularity in the origin. We
use $a=a_B$ and treat $\alpha$ as a fitting parameter. Its values
for different ions were chosen to fit energy levels of doubly ionized 
ions and the same values of $\alpha$ were used for triply ionized 
ions with the same electronic structure. Fine tuning the energies leads
to slightly different values of $\alpha$ for different ions. The
value $\alpha =0.5a_B^3$ was used when data for the fine tuning was
insufficient.

Core polarization modifies both the $4f$ and $5d$ states of an external 
electron; therefore, it would be better to modify both.
However, to obtain correct energy {\em intervals} it is sufficient to
modify either $4f$ or $5d$. Since the $4f$ orbitals are used for both
the self-consistent Hartree-Fock calculations and the CI calculations, it
is simpler to modify the $5d$ states only. As demonstrated previously,
\cite{dzuba-gd},  this procedure works very well for Eu~III 
and Gd~IV.
Therefore, we use it for other rare earth ions as well.

There is one further fitting procedure in the RCI calculations. We scale
Coulomb integrals of second multipolarity by introducing a factor
$f_2=0.8$ before all such integrals. This scaling imitates the 
effect of screening
of the Coulomb interaction of valence electrons by core electrons 
\cite{kozlov}. The value of the factor was chosen to fit energy 
intervals within the $4f^{n-1}5d$ configuration of doubly ionized ions,
where experimental data was available. The same screening
factor was then used for all other ions.

Calculations using the Cowan  atomic structure code \cite{code},
which is also a configuration-interaction method 
(although in its non-relativistic realization), 
are similar to the RCI calculations.  
We use parameters available  in the Cowan code to fit the known energies.
One of these parameters $f_2$ scales the Coulomb integrals
similar to what was done in the
RCI calculations. We use the factor $f_2= 0.85$, which is very close to
the value (0.8) used in the RCI calculations. Another parameter $f_{LS}=0.9$,  
which scales the $L-S$ interaction,  is chosen 
to fit the fine structure of doubly ionized ions.
Finally,  the average energy of the ground state configuration
 was shifted by hand to fit the energy intervals between
the $4f^n$ and the $4f^{n-1}5d$ configurations. Note that this extra energy
shift is due to correlation between core and valence electrons. 
The value of this shift is different for doubly and triply ionized ions.
It follows that the Cowan code alone cannot be used to make predictions of the
energy interval between the $4f^n$ and the $4f^{n-1}5d$ configurations.
On the other hand, the procedure based on formula (\ref{deltaV}) works
well with the same parameters for both ions. Actually, energy
shifts for triply-ionized ions used in the Cowan code were chosen to fit the
results of the RCI calculations. These shifts are 7000 cm$^{-1}$
for Nd~IV, 9000 cm$^{-1}$ for Pm~IV, 10000 cm$^{-1}$ for Sm~IV, and 
12000 cm$^{-1}$ for Eu~IV.

\section{results \label{results}}

In Table~\ref{tab-nd}, we present energies and corresponding
 Land\'{e} $g$-factors for some states of the $4f^{n+1}$ and 
$4f^{n}5d$ configurations
of Nd~IV ($n$ = 2), Pm~IV ($n$ = 3), Sm~IV ($n$ = 4), and Eu~IV ($n$ = 5).
For the $4f^{n}5d$ states we also present lifetimes $\tau$ calculated
by the Cowan code. There are two sets of theoretical energies: RCI 
energies ($E^R$)
and Cowan code energies ($E^C$). For the $4f^{n+1}$ states we also present
experimental energies ($E^N$) from Ref.~\cite{martin}. Note, that there are no 
experimental data for the $4f^n5d$ states of these ions.

The Land\'{e} $g$-factors presented in 
the table are useful for correct identification
of the states. We present $g$-factors as they are calculated by both RCI and
the Cowan code, together with the non-relativistic values, i.e., those
obtained assuming pure
$LS$ coupling. The identification of states is usually easy for low states;
however, states higher in the spectrum are strongly mixed, 
making level identification  difficult, if not impossible. 

Lifetimes of the $4f^{n}5d$ states presented in the table are calculated
with the Cowan code, considering all possible electric dipole transitions 
to lower states of the ground-state configuration  $4f^{n+1}$. Note that
these calculations do not include core polarization and correlation
effects. However, as shown in our calculations for Eu~III
and Gd~IV \cite{dzuba-gd}, these effects are very important. 
For example, they change
the lifetimes of Eu~III and Gd~IV states calculated in the Cowan code
by a factor of 3.6. Similar corrections should be expected for other rare-earth
ions. Thus, we believe that the lifetimes
of the  $4f^{n}5d$ states of Nd~IV, Pm~IV, Sm~IV, and Eu~IV
in the Table~\ref{tab-nd} are too small by a factor of 3 or 4.
 
More detailed theoretical results, together with comparisons
with available experimental data, are presented in the accompanying EPAPS 
document \cite{EPAPS}. Tables~I - IV of \cite{EPAPS} give more detailed 
data for Nd~IV, Pm~IV, Sm~IV, and Eu~IV, respectively. Tables~V - VII
of \cite{EPAPS} give more detailed comparison with experiment
for the ions Pr~III, Nd~III and Sm~III.  Experimental data for these
ions is sufficient for a reliable control of the accuracy of the
calculations. The doubly-ionized ion Pr~III has the same electronic structure 
as triply-ionized Nd~IV. Similarly, the electronic structure of
the pairs Nd~III, PM~IV
and of the pairs Sm~III, Eu~IV  is identical. 
Our calculational scheme can be roughly described 
as choosing parameters for a doubly ionized ion and then running the
programs with the same parameters for the triply-ionized ion with the
same electronic structure. Unfortunately, since there are no experimental data 
for the Pm~III ion, we have no similar control of
calculations for SM~IV. Our confidence in the results
for the Sm~IV ion is based on the fact that the same procedure is used
for all ions and very little adjustment of parameters is needed
to move from one ion to another.

\section{conclusion}

In summary,  systematic  Hartree-Fock  and relativistic
configuration interaction  studies of energy levels and lifetimes of
triply-ionized  rare-earth elements in the $4f^{n+1}$ and $4f^{n}5d$
configurations have been presented. The results of calculations 
using two different sets of computer codes for Nd~IV, Pm~IV, Sm~IV, and 
Eu~IV are in good
agreement with one another and with experiment for the levels of
the ground state configuration $4f^{n+1}$. Our confidence in the accuracy
of energy levels of the  excited configuration
$4f^{n}5d$ is based on similar calculations for Pr~III, Nd~III, and
Sm~III, for which sufficient experimental data is available.

\begin{acknowledgments}
We are thankful to O. P. Sushkov for attracting our attention to the problem.
One of the authors (V.D.) is grateful to the Physics Department of
the University of Notre Dame for the hospitality and support
during his visits in March and June of 2003. The work of W.R.J. was supported
in part by National Science Foundation Grant No.\ PHY-01-39928.
U.I.S. acknowledges partial support by Grant No.\ B516165 from
Lawrence Livermore National Laboratory.
\end{acknowledgments}

\begin{table*}
 \caption{Energies (cm$^{-1}$), Land\'{e} $g$-factors,
 and lifetimes $\tau$ (sec)
calculated by RCI ($E^{\rm R}$) and Cowen code ($E^{\rm C}$)
for Nd~IV ($n$ = 2), Pm~IV ($n$ = 3), Sm~IV ($n$ = 4), and Eu~IV ($n$ = 5).
Comparison with recommended NIST data, $E^{\rm N}$ by
~\protect\citet{martin} is presented.
 \label{tab-nd}}
\begin{ruledtabular}
\begin{tabular}{lrrrrrrlrrrrrr} \multicolumn{4}{c}{} &
\multicolumn{3}{c}{cm$^{-1}$} & \multicolumn{4}{c}{} &
\multicolumn{2}{c}{cm$^{-1}$} &
\multicolumn{1}{c}{sec} \\
\multicolumn{1}{c}{$LSJ$} &
\multicolumn{1}{c}{$g^{\rm NR}$} &
\multicolumn{1}{c}{$g^{\rm C}$} &
\multicolumn{1}{c}{$g^{\rm R}$} &
\multicolumn{1}{c}{$E^{\rm C}$} &
\multicolumn{1}{c}{$E^{\rm R}$} &
\multicolumn{1}{c}{$E^{\rm N}$} &
\multicolumn{1}{c}{$LSJ$} &
\multicolumn{1}{c}{$g^{\rm NR}$} &
\multicolumn{1}{c}{$g^{\rm C}$} &
\multicolumn{1}{c}{$g^{\rm R}$} &
\multicolumn{1}{c}{$E^{\rm C}$} &
\multicolumn{1}{c}{$E^{\rm R}$} &
\multicolumn{1}{c}{$\tau^{\rm C}$}\\
\hline
\multicolumn{5}{c}{$4f^3$ states} &
\multicolumn{4}{c}{Nd~IV ion} &
\multicolumn{5}{c}{$4f^25d$ states} \\
$^4I_{9/2} $& 0.727& 0.732& 0.732&  0   &    0  &  0   &$^2H_{9/2} $& 0.909& 0.807& 0.802& 72100 & 71175 &6.555[-9]\\
$^4I_{11/2}$& 0.965& 0.966& 0.966&  1879&   1945&  1880&$^2H_{11/2}$& 1.091& 0.995& 0.998& 75409 & 74228 &1.747[-8]\\
$^4I_{13/2}$& 1.108& 1.107& 1.107&  3890&   4049&  3860&$          $&      &      &      &       &       &         \\
$^4I_{15/2}$& 1.200& 1.199& 1.199&  5989&   6267&  5910&$^4K_{11/2}$& 0.769& 0.819& 0.810& 72682 & 71864 &6.848[-8]\\
$          $&      &      &      &      &       &      &$^4K_{13/2}$& 0.964& 0.972& 0.971& 75532 & 75172 &7.522[-8]\\
$^4F_{3/2} $& 0.400& 0.426& 0.421& 13294&  12490& 11290&$^4K_{15/2}$& 1.090& 1.090& 1.090& 78311 & 78465 &7.919[-8]\\
$^4F_{5/2} $& 1.029& 1.032& 1.031& 14333&  13545& 12320&$^4K_{17/2}$& 1.176& 1.177& 1.176& 81083 & 81799 &7.752[-8]\\
$^4F_{7/2} $& 1.238& 1.203& 1.224& 15249&  14622& 13280&$          $&      &      &      &       &       &         \\
$^4F_{9/2} $& 1.333& 1.263& 1.183& 16334&  16183& 14570&$^4I_{9/2} $& 0.727& 0.854& 0.860& 74368 & 73977 &9.392[-9]\\
$          $&      &      &      &      &       &      &$^4I_{11/2}$& 0.965& 0.996& 1.012& 74725 & 75234 &6.434[-9]\\
$^2H_{9/2} $& 0.909& 0.962& 1.065& 13272&  14522& 12470&$^4I_{13/2}$& 1.108& 1.105& 1.105& 77187 & 77212 &5.453[-9]\\
$^2H_{11/2}$& 1.091& 1.095& 1.102& 16456&  18142& 15870&$^4I_{15/2}$& 1.200& 1.196& 1.195& 79542 & 79997 &5.400[-9]\\[0.4pc]
$^4S_{3/2} $& 2.000& 1.961& 1.968& 15153&  14452& 13370&$^4G_{5/2} $& 0.571& 0.625& 0.642& 75005 & 74132 &2.371[-8]\\
$          $&      &      &      &      &       &      &$^4G_{7/2} $& 0.984& 0.966& 0.967& 76623 & 76263 &1.536[-8]\\
$^2G_{7/2} $& 0.889& 0.937& 0.956& 18648&  19273& 17100&$^4G_{9/2} $& 1.172& 1.136& 0.998& 78643 & 77442 &1.075[-8]\\
$^2G_{9/2} $& 1.111& 1.133& 1.152& 21376&  21318&      &$^4G_{11/2}$& 1.273& 1.228& 1.030& 80914 & 79265 &1.297[-8]\\
\multicolumn{5}{c}{$4f^4$ states} &
\multicolumn{4}{c}{Pm~IV ion} &
\multicolumn{5}{c}{$4f^35d$ states} \\
$^5I_{4}$&  0.600& 0.604& 0.604&     0&     0&   0 &$^5K_{5}$ & 0.667& 0.690& 0.691& 73812&  73688 &2.410[-8]\\
$^5I_{5}$&  0.900& 0.902& 0.901&  1482&  1550& 1490&$^5K_{6}$ & 0.905& 0.913& 0.913& 75960&  76293 &2.386[-8]\\
$^5I_{6}$&  1.071& 1.071& 1.071&  3110&  3274& 3110&$^5K_{7}$ & 1.054& 1.053& 1.053& 78152&  78952 &2.368[-8]\\
$^5I_{7}$&  1.179& 1.176& 1.176&  4834&  5123& 4820&$^5K_{8}$ & 1.153& 1.149& 1.148& 80377&  81649 &2.313[-8]\\
$^5I_{8}$&  1.250& 1.246& 1.246&  6618&  7063& 6580&$^5K_{9}$ & 1.222& 1.216& 1.215& 82625&  84360 &2.144[-8]\\[0.4pc]
$^5F_{1}$&  0.000& 0.024& 0.019& 14021& 13206&12230&$^5L_{6}$ & 0.714& 0.726& 0.727& 73759&  73372 &2.237[-5]\\
$^5F_{2}$&  1.000& 1.008& 1.004& 14477& 13720&12670&$^5L_{7}$ & 0.911& 0.915& 0.915& 76081&  76202 &4.658[-5]\\
$^5F_{3}$&  1.250& 1.242& 1.245& 15332& 14608&13520&$^5L_{8}$ & 1.042& 1.043& 1.042& 78536&  79188 &8.505[-5]\\
$^5F_{4}$&  1.350& 1.320& 1.337& 16265& 15683&14470&$^5L_{9}$ & 1.133& 1.133& 1.133& 81098&  82300 &9.213[-6]\\
$^5F_{5}$&  1.400& 1.382& 1.389& 17586& 17064&15800&$^5L_{10}$& 1.200& 1.199& 1.198& 83751&  85533 &1.031[-5]\\
\multicolumn{5}{c}{$4f^5$ states} &
\multicolumn{4}{c}{Sm~IV ion} &
\multicolumn{5}{c}{$4f^45d$ states} \\
$^6H_{5/2} $& 0.286& 0.296& 0.295&     0&     0&   0   &$^6L_{11/2}$& 0.615& 0.622&  0.624& 73152& 73945 & 3.057[-5]\\
$^6H_{7/2} $& 0.825& 0.829& 0.829&  1030&  1070&  1080 &$^6L_{13/2}$& 0.851& 0.855&  0.855& 75025& 76264 & 3.632[-5]\\
$^6H_{9/2} $& 1.071& 1.071& 1.071&  2238&  2347&  2290 &$^6L_{15/2}$& 1.004& 1.005&  1.004& 77051& 78765 & 3.529[-5]\\
$^6H_{11/2}$& 1.203& 1.201& 1.201&  3573&  3783&  3610 &$^6L_{17/2}$& 1.108& 1.107&  1.107& 79202& 81411 & 3.559[-5]\\
$^6H_{13/2}$& 1.282& 1.279& 1.279&  4991&  5339&  4990 &$^6L_{19/2}$& 1.183& 1.181&  1.179& 81459& 84189 & 4.711[-5]\\
$^6H_{15/2}$& 1.333& 1.328& 1.329&  6458&  6983&  6470 &$^6L_{21/2}$& 1.238& 1.235&  1.234& 83833& 87119 & 2.998[-5]\\[0.4pc]
$^6F_{1/2} $&-0.667&-0.651&-0.651&  6962&  6584&  6290 &$^6K_{9/2} $& 0.545& 0.560&  0.560& 75414& 76998 & 7.004[-6]\\
$^6F_{3/2} $& 1.067& 1.057& 1.062&  7193&  6886&  6540 &$^6K_{11/2}$& 0.839& 0.845&  0.845& 77095& 79043 & 9.359[-6]\\
$^6F_{5/2} $& 1.314& 1.304& 1.307&  7676&  7457&  7050 &$^6K_{13/2}$& 1.015& 1.014&  1.015& 78892& 81233 & 1.955[-5]\\
$^6F_{7/2} $& 1.397& 1.390& 1.391&  8531&  8375&  7910 &$^6K_{15/2}$& 1.129& 1.123&  1.123& 80757& 83514 & 4.212[-5]\\
$^6F_{9/2} $& 1.434& 1.430& 1.430&  9713&  9614&  9080 &$^6K_{17/2}$& 1.207& 1.196&  1.196& 82654& 85844 & 3.911[-5]\\
$^6F_{11/2}$& 1.455& 1.451& 1.451& 11124& 11097& 10470 &$^6K_{19/2}$& 1.263& 1.247&  1.247& 84544& 88180 & 2.177[-5]\\
\multicolumn{5}{c}{$4f^6$ states} &
\multicolumn{4}{c}{Eu~IV ion} &
\multicolumn{5}{c}{$4f^55d$ states} \\
$^7F_{0}$ & 0.000& 0.000& 0.000&     0&     0&    0 &$^7K_{4}$ & 0.400& 0.410& 0.413&  82484& 82987& 1.609[-4]\\
$^7F_{1}$ & 1.500& 1.500& 1.499&   375&   365&  370 &$^7K_{5}$ & 0.767& 0.771& 0.772&  83854& 84708& 1.346[-4]\\
$^7F_{2}$ & 1.500& 1.499& 1.498&  1035&  1030& 1040 &$^7K_{6}$ & 0.976& 0.977& 0.978&  85399& 86641& 9.429[-5]\\
$^7F_{3}$ & 1.500& 1.498& 1.498&  1887&  1918& 1890 &$^7K_{7}$ & 1.107& 1.106& 1.106&  87086& 88749& 6.362[-5]\\
$^7F_{4}$ & 1.500& 1.498& 1.497&  2861&  2967& 2860 &$^7K_{8}$ & 1.194& 1.192& 1.191&  88897& 91011& 4.438[-5]\\
$^7F_{5}$ & 1.500& 1.496& 1.496&  3908&  4129& 3910 &$^7K_{9}$ & 1.256& 1.252& 1.251&  90828& 93425& 3.288[-5]\\
$^7F_{6}$ & 1.500& 1.495& 1.495&  4992&  5373& 4990 &$^7K_{10}$& 1.300& 1.296& 1.295&  92910& 96033& 3.645[-5]\\[0.4pc]
$^5D_{0}$ & 0.000& 0.000& 0.000& 17064& 21229&17270 &$^7H_{2}$ & 0.000& 0.147& 0.094&  86557& 87999& 1.658[-7]\\
$^5D_{1}$ & 1.500& 1.495& 1.493& 18703& 22890&19030 &$^7H_{3}$ & 0.750& 0.756& 0.715&  87656& 89303& 1.435[-7]\\
$^5D_{2}$ & 1.500& 1.493& 1.491& 21107& 25409&21510 &$^7H_{4}$ & 1.050& 1.045& 1.020&  88957& 90813& 2.314[-7]\\
$^5D_{3}$ & 1.500& 1.489& 1.485& 23980& 28462&24390 &$^7H_{5}$ & 1.200& 1.191& 1.021&  90376& 91830& 6.614[-7]\\
$^5D_{4}$ & 1.500& 1.465& 1.482& 27273& 31966&27640 &$^7H_{6}$ & 1.286& 1.276& 1.143&  91847& 93668& 2.054[-6]\\
          &      &      &      &      &      &      &$^7H_{7}$ & 1.339& 1.329& 1.220&  93348& 95591& 3.764[-6]\\
$^5L_{6}$ & 0.714& 0.730& 0.728& 24294& 27327&      &$^7H_{8}$ & 1.375& 1.362& 1.273&  94863& 97527& 2.180[-5]\\
\end{tabular}
\end{ruledtabular}
\end{table*}


\end{document}